\documentclass[reprint,aip,jcp,superscriptaddress]{revtex4-1}
\usepackage{graphicx}
\usepackage{multirow}
\usepackage{amsmath}
\usepackage{amssymb}
\usepackage{mathtools}
\usepackage{blindtext}
\usepackage[colorlinks,citecolor=blue,urlcolor=black]{hyperref}
\usepackage{setspace}

\def\ben{\begin{equation}}
\def\een{\end{equation}}
\def\mr{\mathbf{r}}

\begin{document}
\title{Energy Level Alignment at Molecule-Metal Interfaces from an Optimally-Tuned 
Range-Separated Hybrid Functional}

\author{Zhen-Fei Liu}
\thanks{These authors contributed equally.}
\affiliation{Molecular Foundry and Materials Sciences Division, Lawrence Berkeley National 
Laboratory, Berkeley, California 94720, USA}
\affiliation{Department of Physics, University of California, Berkeley, California 94720, 
USA}
\author{David A. Egger}
\thanks{These authors contributed equally.}
\affiliation{Department of Materials and Interfaces, Weizmann Institute of Science, 
Rehovoth 76100, Israel}
\author{Sivan Refaely-Abramson}
\affiliation{Department of Materials and Interfaces, Weizmann Institute of Science, 
Rehovoth 76100, Israel}
\author{Leeor Kronik}
\email{leeor.kronik@weizmann.ac.il}
\affiliation{Department of Materials and Interfaces, Weizmann Institute of Science, 
Rehovoth 76100, Israel}
\author{Jeffrey B. Neaton}
\email{jbneaton@lbl.gov}
\affiliation{Molecular Foundry and Materials Sciences Division, Lawrence Berkeley National 
Laboratory, Berkeley, California 94720, USA}
\affiliation{Department of Physics, University of California, Berkeley, California 94720, 
USA}
\affiliation{Kavli Energy Nanosciences Institute at Berkeley, Berkeley, California 94720, 
USA}

\date{\today}

\begin{abstract}
The alignment of the frontier orbital energies of an adsorbed molecule  
with the substrate Fermi level at metal-organic interfaces is a fundamental observable of significant practical 
importance in nanoscience and beyond. Typical density functional theory calculations, especially those 
using local and semi-local functionals, often underestimate level alignment leading to inaccurate
electronic structure and charge transport properties. In this work, we 
develop a new fully self-consistent predictive scheme to accurately compute level alignment at
certain classes of complex heterogeneous molecule-metal interfaces based on optimally-tuned 
range-separated hybrid functionals. Starting from a highly accurate description of the gas-phase 
electronic structure, our method by construction captures important nonlocal surface polarization 
effects via tuning of the long-range screened exchange in a range-separated hybrid in a non-empirical and system-specific 
manner. We implement this functional in a plane-wave code and apply it to several 
physisorbed and chemisorbed molecule-metal interface systems. Our results are in 
quantitative agreement with experiments, both the level 
alignment and work function changes. Our approach constitutes a new practical scheme for accurate and 
efficient calculations of the electronic structure of molecule-metal interfaces. 
\end{abstract}

\maketitle

\section{Introduction}

Interfaces between molecules and metals play a central role in emerging functional devices 
in nanoscience and nanotechnology \cite{I99,KKG03,K07,UK08,HWK09,B09,KUW13,L14,W15,M16}. When a molecule is adsorbed on a metal 
surface, several important physical and chemical phenomena occur. For example, 
an interface dipole forms, altering the work function of the metal surface\cite{L71,I99}; discrete 
molecular frontier orbitals hybridize with extended metallic states, forming molecular resonances;
and substrate screening effects shift orbital energies.
A key physical observable is the energy level alignment between the frontier molecular 
resonance peak positions and the Fermi level, $E_{\rm F}$, of the metal; this alignment can be directly
linked to the energy barrier and efficiency of charge {transfer across} the interface. For example, in molecular 
junctions, highest occupied molecular orbital (HOMO) or the lowest unoccupied molecular 
orbital (LUMO) resonance energies relative to $E_{\rm F}$ are central to determining the zero-bias conductance 
of the junction, as shown in Ref. \citenum{QVCL07}. 
{Molecule-metal bonding spans a range of binding 
energies and degrees of hybridization strengths\cite{B09,L14,W15}, from weak physisorption to strong 
chemisorption; in many cases where covalent or 
molecule-metal interactions associated with significant charge transfer occur, molecular signatures in the 
interface electronic structure are nonetheless observed. For these intermediate cases, molecular resonances can be 
energetically close to or at the metal Fermi level,} which is generally an indication of strong charge transfer between 
the two systems \cite{XDR01} and of significant and complex changes to the interface dipole. {Because level alignment and charge 
transport are intertwined,} its accurate description
is of general importance for understanding, controlling, and predicting functional properties at a variety of interfaces, including systems related to energy conversion and storage.

Experimentally, energy level alignment can be determined by direct 
photoemission spectroscopy for occupied molecular orbitals, and by inverse photoemission 
spectroscopy for unoccupied molecular orbitals.\cite{I99,UK08,HWK09,B09} {Conductance 
measurements} \cite{RZMB97,T06,VKNH06} of molecular junctions probe charge transport properties and, thus, include indirect information about the level 
alignment. However, in principle, distinct binding geometries of molecules in junctions can lead to 
different level alignment \cite{DKKR03} and, hence, {charge transport 
properties} \cite{QKSC09}, and as charge transport measurements are usually ensemble 
averages of multiple geometries, care must be taken in relating conductance to level alignment
in such measurements \cite{QVCL07,QKSC09,DKCV10}. 

First-principles electronic structure calculations that can model 
individual, well-defined geometries provide additional information 
complementary to experiments. {From a formal theory viewpoint}, molecular levels at 
interfaces are quasiparticle energy levels, i.e., {they correspond to} charged excitations. 
A rigorous formalism for quasiparticle energies is many-body perturbation theory (MBPT), 
{which in practice is most often used} in the $GW$ approximation \cite{H65,HL86}, where $G$ is the single-particle Green's function and $W$ is the screened Coulomb interaction. $GW$ has 
been shown to be {an accurate approach} for a {wide} range of molecules \cite{B11,S12,F11,F14,SCSR15,K16} and 
bulk solids \cite{HL85,A01,R05}. However, {several issues have hindered the widespread use 
of $GW$ calculations for molecule-metal interfaces.} Firstly, due to its high computational cost, {even} with today's 
computing power {it is still far challenging} to perform $GW$ calculations of 
several hundred atoms with periodic boundary conditions, as is the case for molecule-metal 
interfaces \cite{LLSS08,KMH14,D14,ST12}. {Moreover, several benchmarking studies, including work on smaller molecule-metal 
systems,} {also showed that results from perturbative $GW$ calculations can be challenging and expensive to 
converge numerically}\cite{SKF06,TDQB11,S12}. {Furthermore,} it is by now well-known that single-shot $GW$ calculations
often depend on the underlying starting point,\cite{R08,B11,BM13} {which adds 
additional complications for efficient $GW$ calculations of large-scale molecule-metal 
interfaces and their functional properties.}

Balancing accuracy and efficiency is crucial for 
calculations of molecule-metal interfaces, and density functional theory (DFT) 
\cite{HK64,KS65} is usually the pragmatic choice for {first-principles calculations at 
relatively moderate computational cost}, provided it is accurate enough. Formally, however, eigenvalues of Kohn-Sham (KS) Hamiltonians are 
only zeroth order approximations to quasiparticle energies\cite{CGB02}. In fact, there is no 
theorem guaranteeing {that they are quantitatively} accurate, {with the important 
exception} of the HOMO energy, for which the ionization potential (IP) theorem holds
\cite{AB85,KSRB12,KK14}. {For the band gap of a semiconductor (or, equivalently, the molecular 
HOMO-LUMO gap), even the exact functional in KS DFT is not necessarily accurate, as it 
lacks a derivative discontinuity\cite{PPLB82} in the exchange-correlation (XC) potential owing to its strictly 
local, multiplicative nature \cite{P83,SS83,K08}.} 

\begin{figure}
\begin{center}
\includegraphics[width=3.5in]{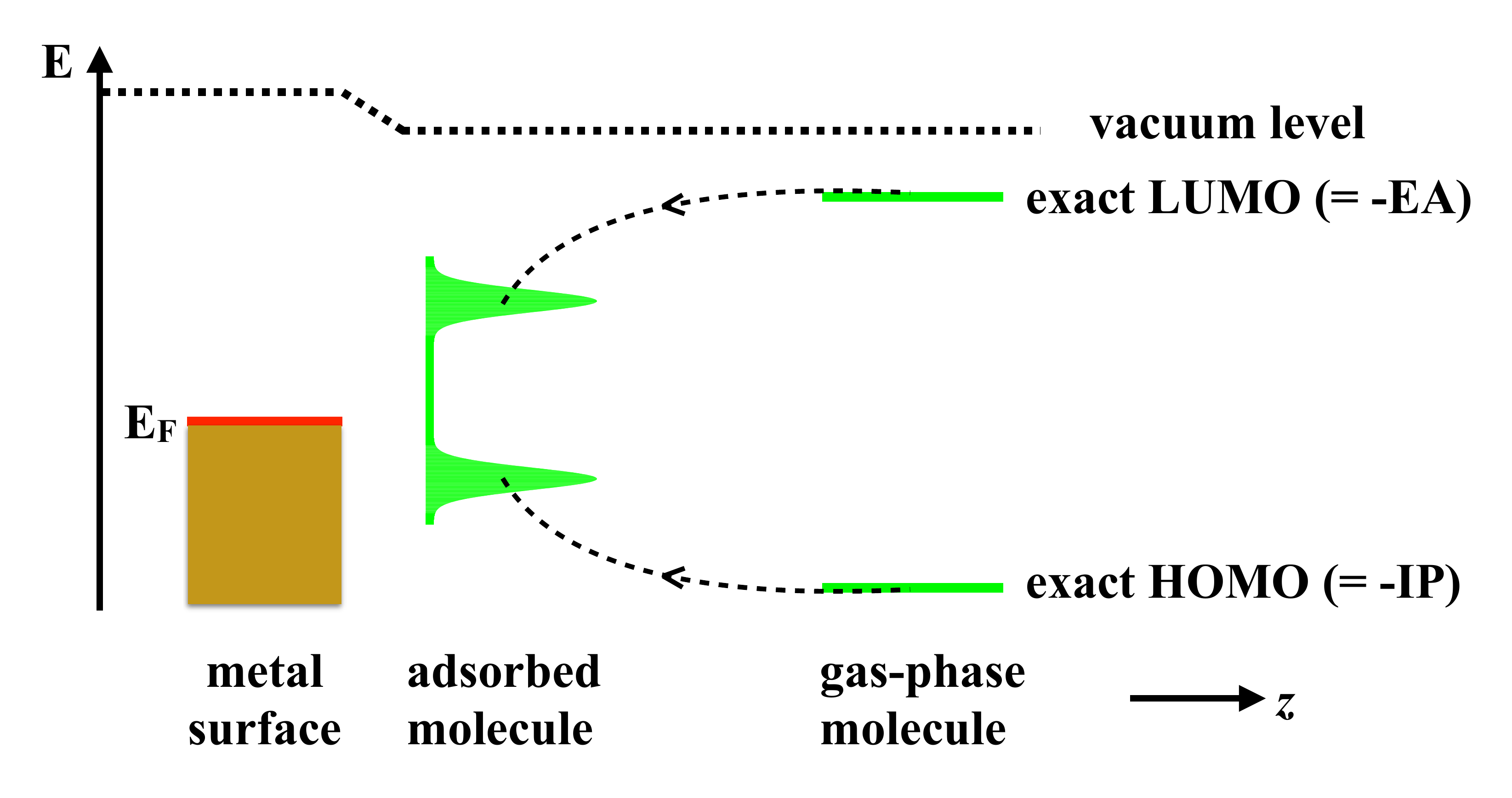}
\caption{Relevant energy levels at a molecule-metal interface. Green lines (shaded areas) {represent} exact 
quasiparticle levels (PDOS of the adsorbed molecule). Note that for the 
case of the adsorbed molecule, broadened molecular resonances are shown. Dashed arrows indicate 
the surface-induced renormalization of the molecular levels, which depends on the molecule-metal distance, $z$. The sign of the vacuum level change at the interface, a result of the interface dipole, is taken here to be negative. For easy visualization, we align the interface-modified vacuum level with that of the gas-phase molecule.}
\label{alignment}
\end{center}
\end{figure}

Fig.\ \ref{alignment} schematically shows representative energy 
levels for the {molecule-metal} interface. The right hand side 
illustrates the situation for an isolated gas-phase molecule, for which there is a 
well-defined quasi{hole level denoted as} 
HOMO, which is equal to the negative of the IP, and a well-defined quasielectron level 
denoted as LUMO, which is equal to the negative of the electron affinity (EA). For the molecule adsorbed on 
the metal surface, in Fig.\ \ref{alignment} we denote resonances {that appear as} peaks in the projected 
density of states (PDOS) - where the DOS has been projected onto the molecular subspace - as HOMO and LUMO, which are now broadened 
due to hybridization. Comparing the molecular energy levels of the gas phase and the 
adsorbed molecule, Fig.\ \ref {alignment} illustrates the important physical phenomenon of surface-induced 
gap renormalization \cite{I73,NHL06,G09,F09}: when the molecule is close to the surface, electrons in the 
metal respond to and screen single-particle excitations in the molecule, i.e., free carriers
in the metal polarize when a hole/electron is added to the molecule and screen the 
Coulomb interactions in the molecule. As a result, when the molecule approaches 
the surface, the HOMO and LUMO energies move closer to $E_{\rm F}$ and the HOMO-LUMO gap is reduced. \cite{I73,NHL06,G09,F09} 
For weakly coupled physisorbed molecules, the energy change of the molecular levels relative to $E_{\rm F}$ 
as a function of the molecule-metal distance has been shown to follow a classical image-like form 
\cite{NHL06}, and renormalization is sometimes referred to as an ``image-charge effect''. 
The effective ``image potential'' is given by $1/[4(z-z_0)]$, where $z$ is the average height of the molecule on the surface and $z_0$ is the image plane position. It is an effective one-body potential resulting from a many-body effect, namely non-local correlation, rather than from a static perturbation to the system.
Fig.\ \ref{alignment} also shows how the vacuum levels are aligned at the interface and for the gas-phase molecule. The vacuum level can increase or decrease at the interface (only one case is shown in Fig.\ \ref{alignment}), due to formation of the interface dipole.

\begin{figure}
\begin{center}
\includegraphics[width=3.5in]{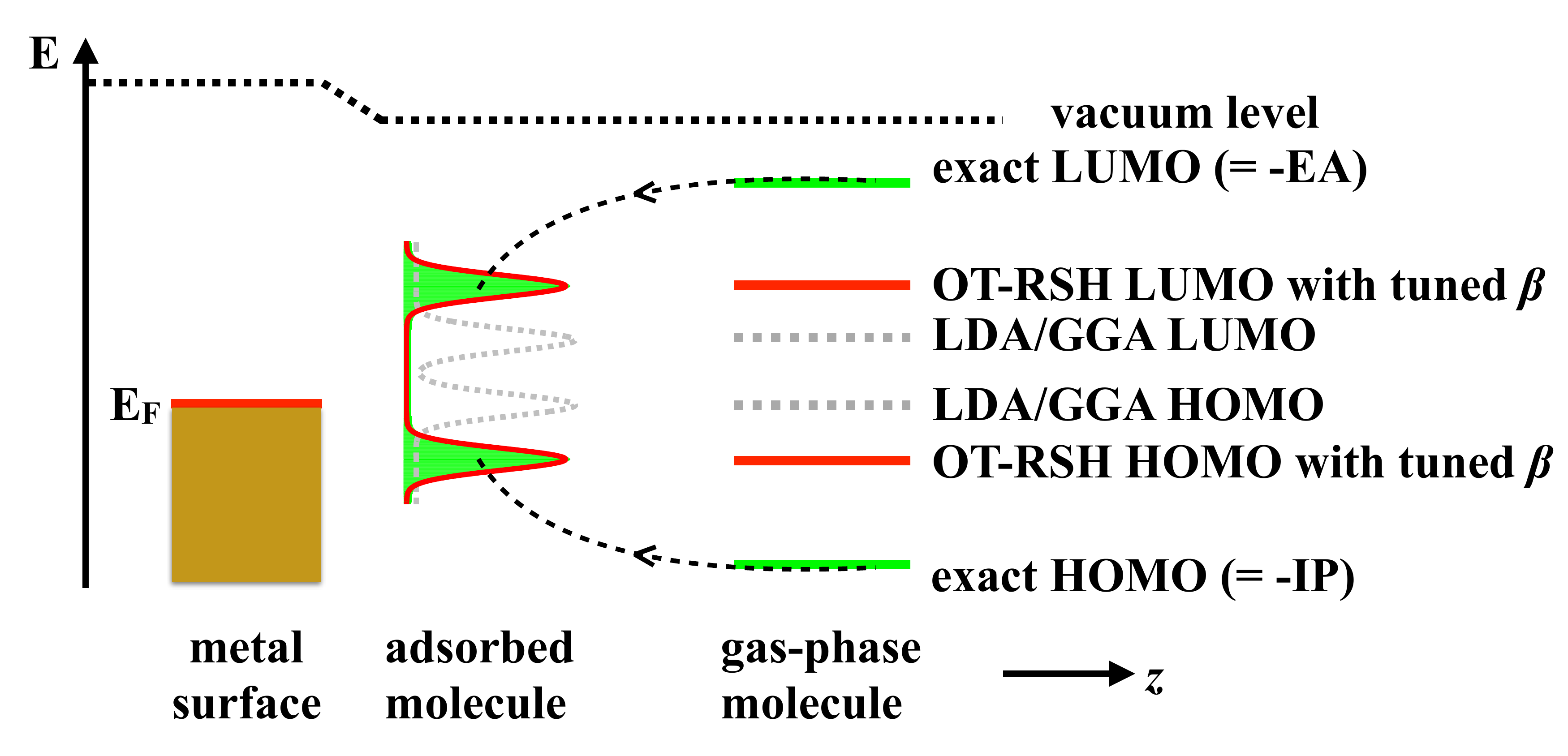}
\caption{Energy levels at a molecule-metal interface. Green lines {represent} exact 
quasiparticle levels, grey dashed lines approximate LDA/GGA levels, and red lines OT-RSH levels as 
computed with the method introduced in this paper (see text for details). Note that for the 
case of the adsorbed molecule, broadened molecular resonances are shown. Dashed arrows indicate 
the surface-induced renormalization of the molecular levels, which depends on the molecule-metal distance, $z$. The sign of the vacuum level change at the interface, a result of the interface dipole, is taken here to be negative. For easy visualization, we align the interface-modified vacuum level with that of the gas-phase molecule.}
\label{rshcartoon}
\end{center}
\end{figure}

Results from calculations 
using the local density approximation (LDA) \cite{KS65} and {typical} generalized 
gradient approximations (GGAs) in the Kohn-Sham (KS) DFT scheme, while often accurate 
for total energy or charge density related properties of molecule-metal interfaces such as 
work-function changes\cite{D05,M08,R09,TRHR10,HAMR13,H14,A16}, 
typically strongly underestimate 
the band gap of molecules and 
semiconductors\cite{HL86}. We {illustrate this} with grey dashed 
lines in Fig.\ \ref {rshcartoon},  namely that LDA/GGA misplace the 
HOMO and LUMO levels in the gas phase, such that the gap is underestimated. 
Moreover, in typical LDA/GGA calculations of molecule-metal interfaces, one 
finds that the {HOMO-LUMO} gap for the adsorbed 
molecule remains virtually the same as in the gas phase. In other words, {LDA and GGA 
results} for the level alignment are not sensitive to {changes in screening 
environment} and fail to capture the gap renormalization effect. This is because LDA and GGAs consist of 
local correlation only whereas renormalization is a nonlocal effect 
\cite{I73,NHL06,G09,F09}. {As we discuss below, similar findings hold for conventional 
global and screened hybrid functionals.} 

The high computational cost {and convergence challenges associated with} $GW$, together with the {formal and practical issues 
of popular XC functionals in predicting level alignments, led to the development of several correction 
schemes following various directions, see Refs. \citenum{NHL06,FOV09,G09,R10,G11,SRPS13,HAMR13,Y13,
M14,E15,Mo16,RMS16,MLNW16} for examples. Specifically 
for physisorbed systems, one can assume weak coupling between the 
molecule and the metal, and attempt to
shift the PDOS peaks in a non-self-consistent, adjustable-parameter-free \emph{a posteriori} manner. One example for such a non-self-consistent 
correction scheme is the DFT+$\Sigma$ {approach}\cite{NHL06}, where $\Sigma$ denotes a two-step correction 
to KS eigenvalues: (1) it uses a gas-phase correction, equal to the difference between the gas-phase LDA/GGA HOMO or LUMO {energy} 
and {a value obtained from a more accurate approach (e.g., $GW$) that is closer to} the 
exact {quasiparticle energy (cf. Figs. \ref{alignment} and \ref{rshcartoon})}; and (2) a correction {accounting for the} surface 
polarization responsible for gap renormalization, which is often approximated using a classical 
image-charge model \cite{NHL06}, $1/[4(z-z_0)]$, where $z$ is the average distance between 
the adsorbed molecule and the surface, and $z_0$ is the image-plane position. We note that recently we have augmented this approach by 
using a non-classical DFT-determined image plane of the metallic surface to compute the surface polarization term, rather than 
the classical DFT-derived image plane\cite{ELNK15}. The DFT+$\Sigma$ method has been used to successfully 
predict and explain level alignment at {physisorbed} interfaces \cite{ELNK15} and charge 
transport in specific molecular junctions where the above-mentioned weak-coupling assumption is reasonable \cite{QVCL07}.}

{Of major interest are approximate $GW$-inspired theoretical approaches that can go beyond the weak-coupling assumption to also treat molecules 
chemisorbed on metals, i.e., when a significant amount of charge is transferred between the two subsystems. Examples include systems 
for which molecular levels are pinned at $E_{\rm F}$ \cite{VOPO04,HDSG13}, or where significant covalent interactions lead to strong hybridization between molecular and substrate orbitals \cite{HRBZ06}, 
such as the Au-benzenedithiol-Au junction \cite{SRPR14,RRO15}. 
As mentioned above, this situation can lead to the
molecular resonances being energetically close to the Fermi level, which implies a significant charge transfer. Clearly, 
the amount of charge transfer will determine 
the distribution of the charge density of the interface system, which means that a correction scheme for level alignment requires self-consistency.  
This notion also implies that in case of chemisorption an incorrect description of level alignment can yield errors in the 
density and, at least in principle, result in incorrect interface dipoles,\cite{HAMR13} work functions, and total energies 
of molecule-metal systems. Therefore, {we currently seek} a theoretical method that can accurately and efficiently 
characterize the interface electronic structure in a self-consistent manner, and thus is applicable to physisorbed and, in particular, 
also to more strongly bound and partially chemisorbed molecule-metal systems. With such an approach in hand, one is able to improve 
the prediction of relevant interface properties beyond level alignment, including work function 
changes and PDOS lineshapes.}

{Here, we present a new method aiming in the vein of} reliable and self-consistent 
determination of the level alignment {at molecule-metal interfaces}. Our {conceptual 
framework is} the generalized Kohn-Sham (GKS) scheme \cite{SGVM96}, {which is still a 
DFT approach but in contrast to LDA/GGA builds on an} effective potential that is nonlocal. 
{Hybrid functionals, in which} (semi)local functionals are mixed with a 
fraction of Fock exchange, {are a special case of GKS DFT.} {Popular hybrid functionals} 
were shown to greatly improve GGA results for both total energy related properties, such as 
thermochemistry \cite{B93}, as well as quasiparticle energy levels, such as band gaps 
\cite{HPSM05}. {However, conventional hybrid functionals do not in general 
remedy the energy level alignment problem at interfaces
\cite{BTNK11,ELNK15}. Notably, conventional hybrids cannot capture the renormalization of the HOMO-LUMO gap. A relatively recent and particularly promising class are} optimally-tuned 
range-separated hybrid (OT-RSH) functionals \cite{B10,KSRB12}, in which parameters of 
the functional are tuned {per system} to satisfy important physical conditions, {without 
recourse to empirical data or any fitting}. OT-RSH was shown to yield accurate frontier 
orbital energy levels, {outer-valence electron spectra, and HOMO-LUMO gaps} of gas-phase 
molecules\cite{S10, RA11, K11, RSGA12, EWRS14, P14, TRNK14, KB14, A14} and molecular crystals 
\cite{RSJB13,KN16}, as well as transport properties \cite{LWYA14,YFHF16}. In this work, we extend the 
applicability of this functional to heterogeneous molecule-metal interfaces {by proposing a 
route for} a judicious choice of {optimal} parameters {in the OT-RSH functional}. {This 
yields a fully self-consistent} OT-RSH scheme {which is applicable to both physisorbed 
and chemisorbed molecule-metal systems and thus extends methods reliant on the weak 
coupling limit.} Using it for several prototypical test cases, we achieve quantitative agreement 
with experiments for level alignment {and work function changes} including 
those featuring charge transfer and stronger bonding.

The outline of this paper is {as follows:} In Sec. \ref{sec2}, we briefly review the OT-RSH 
functional as applied to gas-phase molecules and molecular crystals, and then present our 
approach for treating molecule-metal interfaces. In Sec. \ref{sec3}, we apply the method to 
six {molecule-metal interfaces} that have {been well-studied} in both theory 
and experiment, including two {systems} where charge transfer and Fermi level pinning 
occur. In Sec. \ref{sec4}, we discuss limitations of the proposed method and {outline 
remaining challenges, which is followed by our conclusions} in Sec. \ref{sec5}.

\section{Methodology}
\label{sec2}

\subsection{OT-RSH for gas-phase molecules and molecular crystals}
In RSH functionals, the Coulomb operator is decomposed {into short-range 
and long-range components \cite{TCS04}. In this approach, proposed by Yanai et al.}\cite{Y04}, the decomposition takes the form
\ben
\frac{1}{\left| \mr-\mr'\right|}=\frac{\alpha+\beta \mbox{erf}\left(\gamma \left| \mr-
\mr'\right|\right)}{\left| \mr-\mr'\right|}+\frac{1-\left[\alpha+\beta \mbox{erf}
\left(\gamma \left| \mr-\mr'\right|\right)\right]}{\left| \mr-\mr'\right|},
\label{otrsh}
\een
where $\alpha,\beta,$ and $\gamma$ are parameters and $\mbox{erf}\left(\cdot\right)$ is the error function. 
We note that this partition is not unique, but the choice of the error function is computationally convenient. 
Here, we treat the first term 
using nonlocal Fock exchange and the second using semi-local exchange. 
The exchange-correlation energy {can then be} written as
\ben
\begin{split}
E_{\rm XC} = & \, \alpha E_{\rm X, SR}^{\rm EXX} + \left(1-\alpha\right) E_{\rm X, SR}^{\rm 
GGA} \\
& + \left(\alpha+\beta\right) E_{\rm X, LR}^{\rm EXX} + \left(1-\alpha-\beta\right) E_{\rm 
X, LR}^{\rm GGA} \\
& + E_{\rm C}^{\rm GGA}, \\
\end{split}
\label{eng}
\een
where the subscript X (C) {denotes} exchange (correlation), SR (LR) denotes the short-range 
(long-range) exchange, and the superscript EXX (GGA) reflects whether the corresponding energy 
component is treated using Fock exchange (GGA {exchange} or correlation). In this work, we 
follow Ref. \citenum{RSGA12} and use PBE (Perdew-Burke-Ernzerhof) \cite{PBE96} for the {GGA 
exchange ($\omega$PBE \cite{HJS08} for the SR part) and correlation components.} 

For an isolated gas-phase molecule, $\alpha$ is often chosen {to be} 0.2,\cite{RSGA12,EWRS14,LRPR14} and then $\beta$ 
is chosen as $1-\alpha=0.8$ to ensure the correct asymptotic potential\cite{AB85} via enforcing 
full long-range Fock exchange. {Here}, we will use {the notation} $\beta_0=0.8$ to denote the $\beta$ 
value appropriate for isolated gas-phase molecules. $\gamma$ is the {range-separation} parameter and 
governs the separation between SR and LR. {In the OT-RSH approach,\cite{S10,B10,KSRB12} the 
remaining parameter} $\gamma$ is tuned {separately for each system} by minimizing
\ben
J(\gamma)=\left|\varepsilon^{\gamma}(N)+\mbox{IP}^{\gamma}(N)\right|^2+\left|
\varepsilon^{\gamma}(N+1)+\mbox{IP}^{\gamma}(N+1)\right|^2,
\label{tune}
\een
where $\varepsilon$ {are} GKS HOMO eigenvalues, the IP is calculated using total energy 
differences between neutral and charged systems, and $N$ ($N+1$) indicates the neutral molecule (anion). 

Ref. \citenum{RSJB13} generalized the OT-RSH functional to {the case of} molecular 
crystals. $\alpha$ and $\gamma$ {were} chosen according to the gas-phase values, but $\beta
$ was {adapted} to {account for changes} in the long-range Coulomb screening due to {the 
presence of the} other molecules in the crystal {and the difference in dielectric 
environment} compared to an isolated molecule. From Eq. \eqref{eng}, we {see} that $\alpha+
\beta$ governs the fraction of long-range Fock exchange. For a molecular crystal with an average 
dielectric constant $\epsilon$, Ref. \citenum{RSJB13} proposed {to use} $\alpha+
\beta=1/\epsilon<1$. This choice of $\beta$ {was shown} to correctly describe the gap 
renormalization in molecular crystals compared to a single gas-phase molecule 
and optical absorption\cite{R15}.

\subsection{OT-RSH for molecule-metal interfaces}

We first note that for a fixed, $z$-independent choice of $\alpha,\beta,$ and $\gamma$, 
the OT-RSH functional \emph{cannot} capture the $1/[4(z-z_0)]$ behavior of the image-charge 
effect and gap renormalization of the adsorbate in the weak-coupling physisorbed limit at a metal surface for all $z$. To 
understand why, consider that in Ref. \citenum{NHL06}, it was shown that the physical 
origin of the gap renormalization is the change in the screened Coulomb interaction, $
\Delta W$, between the isolated molecule and the adsorbate. From a GKS viewpoint, this 
long-range correlation effect would require a \emph{change} in the amount of long-range 
screened Fock exchange as a function of molecule-metal distance, which Eq. \eqref{eng} with 
fixed parameters lacks. In fact, this is also the case for any standard local, semi-local, 
or hybrid functional, as was shown in Ref. \citenum{BTNK11} using the PBE and HSE 
(Heyd-Scuseria-Ernzerhof) \cite{HSE03} functionals as examples.

{As described above,} in Ref. \citenum{RSJB13}, $\beta$ was tuned from $\beta_0$
to $1/\epsilon-\alpha$ in order to capture the change in Coulomb screening between 
an isolated molecule and a molecular crystal. Inspired by this idea, and given the fact that
the image-charge effect is a long-range effect, we propose {to account for 
the change in long-range Fock exchange} at a molecule-metal interface by 
{insisting on} $\alpha+\beta<1$ for the interface and tuning $\beta$ so as to capture the renormalization of the orbital resonance energies at the level of an image-charge model. {For $\alpha$ and $\gamma$, it is clear 
that in the limit of very weak physisorption the optimal value obtained in the gas phase 
is maintained on the surface as the molecular density remains unchanged. 
For chemisorbed systems with exchange of charge between molecule and metal, of course 
this does not hold anymore, but as we show below the optimal value for 
$\gamma$ is virtually unaffected by small charge density rearrangements.} 
Thus, for each molecule-metal system we choose $\alpha=0.2$ and $\gamma$ based on Eq. \eqref{tune} 
as obtained for the gas-phase molecule. The problem left is how to choose the 
optimal $\beta$ for the interface such that the renormalization induced by the surface is 
accounted for (see red lines of Fig.\ \ref{rshcartoon}). Here, we choose $\beta$
such that the orbital energies 
renormalize properly according to $1/[4(z-z_0)]$. To avoid a complicated functional form 
with explicit treatment of this effect,  
we here use a DFT-based image-charge model: {it allows for determining a non-classical 
image-plane position for each type of surface in a unique way} 
\cite{ELNK15}, from which we determine the amount of orbital renormalization. If we keep 
$\beta$ as a 
simple scalar parameter, it {should} implicitly depend on $z$ due to the $z$-dependence
in the image-charge model. Furthermore, we make the assumption that the orbital 
energies of the isolated molecule and the PDOS peaks of the adsorbate change by the same 
amount when $\beta$ is varied from $\beta_0$ to $\beta<1-\alpha$, while keeping $\alpha$ and 
$\gamma$ at their gas-phase values. Therefore we can tune $\beta$ in the gas phase, which 
reduces the computational cost, and choose $\beta$ such that the neutral gas-phase HOMO 
changes by an amount equivalent to the image-charge energy, such that
\ben
\varepsilon_{N}^{\gamma}(\beta)-\varepsilon_{N}^{\gamma}(\beta_0)=P,
\label{chooseb}
\een
where subscript $N$ {denotes} the neutral gas-phase system and $P=1/[4(z-z_0)]$ is the 
image-charge energy determined as in Ref. \citenum{ELNK15}. 
{We note that the substrate provides significant screening, and the $\beta$ defined in this way can become negative. Considering 
that $\beta=-\alpha$ corresponds to the limit of fully screened LR Fock exchange, we choose 
this value as our lower limit (i.e., $\beta\geq-0.2$ when $\alpha$ is chosen to be 0.2).}
From our experience, $\varepsilon_{N}^{\gamma}(\beta)$ changes linearly {with} $\beta$ for 
all the molecules we 
studied. A cartoon for our $\beta$-tuning {scheme} is 
shown in Fig.\ \ref{rshcartoon}, where the red line on the right indicates the gas-phase HOMO 
calculated using the tuned $\beta$. 

Our tuning {procedure,} for a given molecule-metal interface system with $z$ obtained from a prior DFT calculation, {can be summarized as} follows:
\begin{enumerate}
\item
Perform a standard gas-phase OT-RSH calculation for the molecule, {i.e.}, use $
\alpha=0.2, \beta_0=1-\alpha=0.8$, and determine $\gamma$ as in Eq. \eqref{tune};
\item
For the metal slab, compute the image-plane {position,} $z_0$, by matching the long-range XC potential from a local or semi-local functional {(PBE in this work)} to an image potential.
{To be specific, first compute the $xy$-averaged PBE XC potential for a given metal slab, $V^{\rm PBE}_{\rm XC}(z)$, where $z$ is the variable describing the distance away from the metal surface, then tune the parameter $z_0$ such that the two curves $-1/[4(z-z_0)]$ and $V^{\rm PBE}_{\rm XC}(z)$ have a common tangent point (see Refs.\ \citenum{EH89,LLG09,ELNK15} for details)};
\item
Compute the {approximate polarization induced by the surface with a classical image-charge 
model} {by evaluating the expression} $P=1/[4(z-z_0)]$ (in a.u.), where $z$ is determined from the geometry {of the interface and} is the average distance between the 
molecule and the top layer of the metal slab along the surface normal;
\item
Tune the optimal $\beta$ value according to Eq. \eqref{chooseb}, using a series of OT-RSH 
calculations of the gas-phase molecule, with varying $\beta$ and fixed $\alpha$ and $\gamma
$;
\item
Compute the electronic structure of the molecule-metal interface 
in a self-consistent manner using this OT-RSH functional with the above-determined $\alpha$, 
$\beta$, and $\gamma$ (in general, each molecule-metal system has its own set of {optimal} parameters).
\end{enumerate}
In this work, we stop at Step 5. However in principle, one could re-optimize the geometry of the molecule-metal interface, obtain a new $z$, and iterate the tuning procedure to self-consistency.

\subsection{Technical details}
\label{tech}
All OT-RSH calculations of molecule-metal interfaces are performed using a modified 
version of Quantum ESPRESSO \cite{QE2009} v. 5.2.0. 
We implemented long-range screened Fock exchange based on the existing 
subroutines of short-range screened Fock exchange and hybrid functionals, using norm-conserving 
pseudopotentials. This is realized via the Fourier transform of the first term in Eq. 
\eqref{otrsh}, that is
\ben
\frac{\alpha+\beta \mbox{erf}\left(\gamma \left| \mr-\mr'\right|\right)}{\left| \mr-
\mr'\right|} \xRightarrow{\mbox{F.T.}} \frac{4\pi}{q^2}\left[\alpha + \beta \mbox{exp}
\left(-\frac{q^2}{4\gamma^2}\right)\right].
\een
The Gygi-Baldereschi approach \cite{GB86} is used to treat the Coulomb potential divergence 
at small $q$ vectors, as already implemented in Quantum ESPRESSO. 
A 55 Ry energy cutoff is used for every system. Fock exchange is 
evaluated using a reduced k-mesh in our calculations; {for smaller systems, we verified 
convergence with respect to the k-mesh}. The k-mesh used for each system is specified in Sec. \ref{sec3}.

{In the current implementation of hybrid functionals in Quantum ESPRESSO, there is an 
inner self-consistency loop which solves the GKS equation with a fixed Fock operator, 
and an outer self-consistency loop which updates the Fock operator with new GKS orbitals.} 
{Unless noted otherwise, all} results shown in this work are from ``perturbative hybrid'' calculations {(which are still self-consistent in a certain sense, see below)}, in {which} the Fock operator is constructed only once, using pre-converged PBE 
orbitals, and the resulting GKS equation {is then solved {self-consistently}} to obtain eigenvalues and 
orbitals without updating the Fock operator again. {The resulting eigenvalues, eigenvectors, and charge density are already different from their PBE counterparts and this self-consistent solution of GKS equation corrects the large errors of PBE in PDOS calculations, in terms of both lineshape and peak positions.} Our tests show that this 
``perturbative hybrid'' {approach} yields almost identical PDOS around the Fermi level 
compared to {fully} self-consistent hybrid calculations with both inner and outer loops, but is 
much more efficient {computationally} because we omit the outer self-consistency loop. 

It is {well-}known that for transition metals, core-valence interactions may have non-negligible 
numerical effects in $GW$ \cite{MOD02} and exact exchange-based \cite{SDA05} calculations 
of semiconductor band structures. In our tests, the semicore $sp$-states of transition 
metal atoms are found to affect the level alignment by about 0.3 eV. However, {explicit 
inclusion of} the semicore $sp$-states significantly increases the total number of electrons 
as well as the required energy cutoff. In order to reduce 
computational costs, we use a pseudopotential (see Ref. \citenum{ppnote} for details)
with semicore $sp$-states (e.g., 4s$^2$4p$^6$4d$^{10}$5s$^1$ for Ag) 
in the valence for the top layer of the transition metal 
substrate that is closest to the adsorbed molecule, and a pseudopotential without semicore $sp$-states 
(e.g., 4d$^{10}$5s$^1$ for Ag) for the reminder of the metal slab {(see Ref. \citenum{mixpp} for a note regarding the accuracy of mixing pseudopotentials).} In addition, we find that 
although total energy convergence requires a cutoff of about 200 Ry with explicit 
semicore $sp$-states, the eigenvalues and energy level alignments typically converge at a 
much lower energy cutoff. The 55 Ry cutoff we use is sufficient for the level alignment 
of all the systems studied in this work.

The geometries of the molecule-metal interfaces are either adapted from {the} literature or 
relaxed using dispersion-corrected XC functionals (see the results section for details), as 
implemented in VASP \cite{vasp} with the Projector Augmented Wave (PAW) method 
\cite{B94,KJ99}. We then use the relaxed molecular adsorbate geometry
in our gas-phase tuning {procedures}. {Note that this 
can result in slightly different ($<0.1$ eV) HOMO values compared with those obtained from using the molecular geometry 
optimized in the gas phase.} Gas-phase tuning based on 
Eq. \eqref{tune} is performed using QChem \cite{Qchem4} with a cc-pVTZ basis set, and gas-phase 
tuning based on Eq. \eqref{chooseb} is performed using NWChem \cite{nwchem} with the same basis 
{set}. 

\section{Results}
\label{sec3}
{To demonstrate the success of our approach,} we report OT-RSH calculations of six molecule-metal 
interfaces. We divide them into 
two categories: {systems} without significant charge transfer (i.e., physisorbed); and 
systems with non-negligible charge transfer (i.e., partially chemisorbed). The latter systems feature 
Fermi level pinning in the PDOS and a significant overlap between the LUMO resonance peak 
and $E_{\rm F}$. {As discussed in the introduction, the OT-RSH method has repeatedly been found to be highly accurate for gas-phase molecules. Therefore, here we do not dwell on gas-phase results but instead focus on the molecule-surface interaction.}

{As discussed above, correction schemes that assume weak coupling and do not update the charge density, 
such as the} DFT+$\Sigma$ approach, are {known to perform very} well for interfaces in the 
first category \cite{ELNK15}. But a self-consistent method, such as the one 
presented in this paper, is needed for accurate calculations of the second category. {We note that there are 
specific cases of chemisorption for which charge transfer involves not the frontier, but energetically deeper lying 
molecular resonances. In such cases, our non-self-consistent scheme may work equally well \cite{Y13}.} For each system, we 
compare PBE and OT-RSH results with {literature results from} 
ultraviolet photoemission spectroscopy (UPS) measurements {for} both {energy} level 
alignment {(taken as position of the peak maximum)} and work function changes. {For the calculated PDOS, a 0.01 Ry broadening is applied to every system, except for the one in Sec. \ref{sec:antse}, where a 0.02 Ry broadening is applied in order to match the experimental PDOS width.}

\subsection{Systems without significant charge transfer}
\label{sec41}
In this section, we present {results for} four systems. To begin, we consider benzene adsorbed on Al(111) and 
3,4,9,10-perylene-tetracarboxylic-dianhydride (PTCDA) adsorbed on Au(111), two weakly 
coupled, physisorbed interfaces {for which we expect} the standard{, perturbative} DFT+$
\Sigma$ {approach} to {perform} well.\cite{ELNK15} These systems {serve as a proof of concept for the 
approach} proposed in this paper. We then consider 1,4-benzenediamine (BDA) adsorbed on 
Au(111), {where charge transfer is negligible but hybridization between molecular orbitals and 
Au $d$-states} is stronger than for the first two systems. In this case, a self-consistent 
approach such as the one presented {here} {may} change the GGA PDOS lineshape, while a 
post-processing approach, such as the standard DFT+$\Sigma$, does not. Finally, we consider Anthracene-2-selenolate 
(AntSe) adsorbed on Au(111), an even more complicated {case} as the Se-H 
bond is replaced by a Se-Au bond upon adsorption, i.e., a new covalent bond is formed.

\subsubsection{Benzene on Al(111)}
Experimental evidence for a rather weak interaction between benzene and Al(111) was 
given in Ref. \citenum{DMBN00}, where UPS in ultrahigh vacuum was used to determine the 
{energy }level alignment. In our calculations of this system, we use a $4\times4$ surface unit cell with 4 layers of Al atoms 
to represent the slab, and employ a $4\times4\times1$ k-mesh ($2\times2\times1$ for the 
Fock exchange contribution). {To determine the geometry of this interface, we perform dispersion-corrected 
DFT calculations using the PBE-TS approach (i.e., PBE augmented by the 
Tkatchenko-Scheffler scheme\cite{T09}) including self-consistent screening \cite{TDCS12} to 
capture the impact of the metallic screening on dispersive interactions for a set of fixed 
molecular geometries at various distances}. The lowest total energy is achieved for the 
molecule lying flat at 3.24 \AA~above the Al(111) surface on a bridge site.

\begin{figure}
\begin{center}
\includegraphics[width=3.5in]{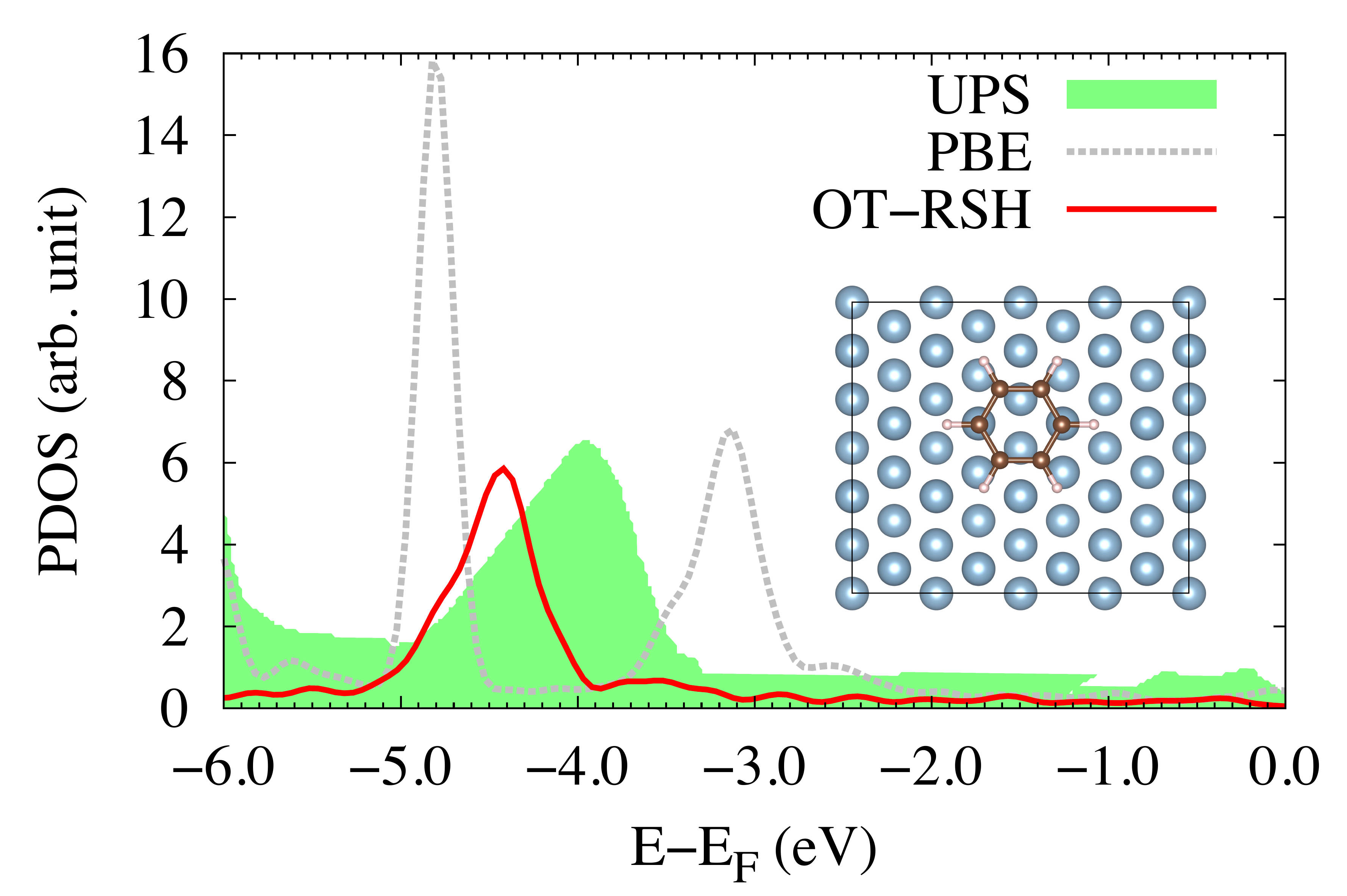}
\caption{PDOS of benzene adsorbed on Al(111), whose structure is shown in the inset, 
as obtained using PBE {(grey dashed line)} and OT-RSH {(red line)}, compared with {results from UPS measurements} 
(green) adapted from Ref. \citenum{DMBN00}.}
\label{f:bz}
\end{center}
\end{figure}

Following the tuning protocol for interfaces described in Sec. \ref{sec2}, Eq. \eqref{tune} yields $\gamma=0.24$ bohr$^{-1}$ and 
{a tuned} HOMO energy of -9.4 eV for benzene. For Al(111), the image plane 
is {determined to be} 1.1 
\AA~above the surface and, {using the optimized distance of our benzene adsorbate}, the image-charge 
energy is then 1.7 eV. In order to incorporate the polarization energy due to the metal 
surface, we tune $\beta$ such that the gas-phase HOMO increases by 1.7 eV, based on Eq. 
\eqref{chooseb}. This results in {an optimal $\beta$ value of $0.20$}. 

Fig.\ \ref{f:bz} shows the experimental {data for level alignment} (as adapted from Ref. 
\citenum{DMBN00}) and {our theoretical} results for the PDOS. OT-RSH yields a HOMO 
{resonance} at 4.4 eV below $E_{\rm F}$, in {much} better agreement with experiment (4.0 eV) than 
PBE (see Fig.\ \ref{f:bz}), which predicts the resonance at 3.1 eV below $E_{\rm F}$. 
{However,} PBE and OT-RSH yield {essentially the same} work function change of -0.3 eV {and the same work function of 3.8 eV}, 
suggesting our new OT-RSH scheme leads to results on par with LDA/GGA for predicting accurate work 
functions and interface dipoles.\cite{D05,M08,R09,TRHR10,HAMR13,H14,A16} We note that the standard, non-self-consistent DFT+$
\Sigma$ yields a HOMO resonance within 0.1 eV difference from OT-RSH result, as expected 
for such a weakly coupled interface. We also {note that when} the vdW-DF2 functional 
\cite{LMKL10} {is used} to relax the coordinates of the molecule and the top layer of Al(111), the benzene molecule is found at about 
3.5\AA~above the Al(111) surface, {which results in a} level alignment of 4.2 eV based on 
the OT-RSH scheme.

\subsubsection{PTCDA on Au(111)}
PTCDA {molecules adsorbed on noble metals have been very well-studied (see Ref. 
\citenum{T07} for a detailed review). It is known that the interaction between 
PTCDA and} Au(111) is rather weak \cite{H04,HBLS07,T07,DGSB08,R09,RLZS12}. For consistency 
with previous work \cite{ELNK15}, our calculations use the same {geometry} as in Ref. \citenum{RLZS12},  
where {dispersion-corrected functionals yields excellent agreement with experiment  
for the structure, and} 3 layers of Au(111) {are used} {to represent the slab;} a $2\times2\times1$ 
k-mesh ($1\times1\times1$ for the Fock exchange contribution) {is employed}. 

{Following our OT-RSH tuning procedure,} in the gas-phase, the optimal $\gamma$ is 
0.16 bohr$^{-1}$ and the OT-RSH HOMO is -8.2 eV. For Au(111), the image plane is 0.9 
\AA~above the top surface, as determined in Ref. \citenum{ELNK15}. The average distance 
between the molecule and the top layer {of the} surface is 3.18 \AA, which yields an 
image-charge renormalization energy of 1.6 eV. We find {that $\beta=-0.10$ is required to modify the long-range 
Fock exchange} and {adjust the} gas-phase HOMO {energy} by 1.6 eV, according to Eq. 
\eqref{chooseb}. 

\begin{figure}
\begin{center}
\includegraphics[width=3.5in]{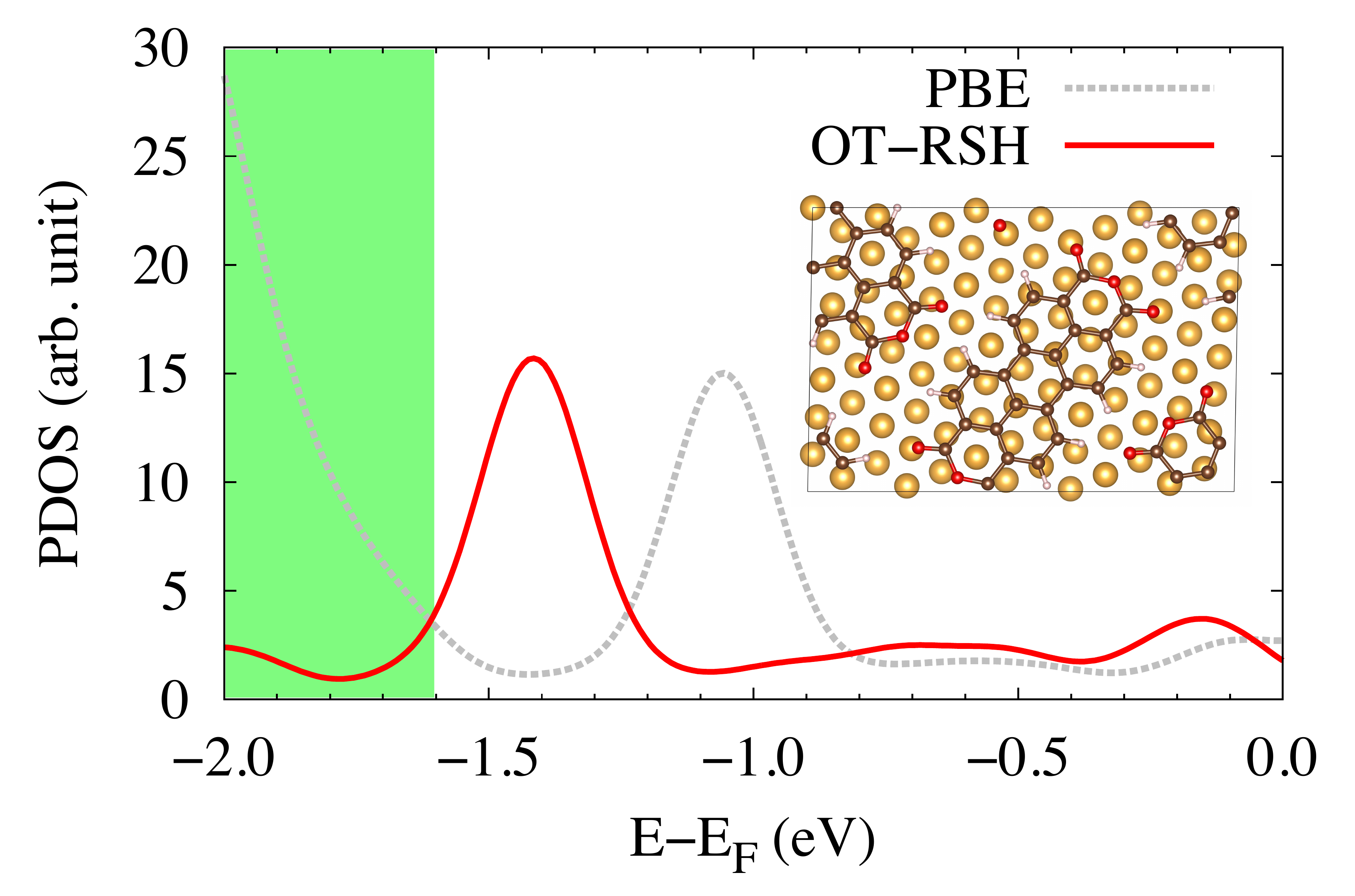}
\caption{PDOS of PTCDA adsorbed on Au(111), whose structure is shown in the inset, as obtained 
using PBE 
{(grey dashed line)} and OT-RSH {(red line)}, compared {with results from UPS measurements 
(shown as green shaded area)} from Ref. \citenum{DGSB08}.}
\label{f:ptcdau}
\end{center}
\end{figure}

Fig.\ \ref{f:ptcdau} shows {our} results: OT-RSH yields 1.4 eV {for the energy level 
alignment of the HOMO resonance relative to $E_{\rm F}$}, which is reasonably close to the experimental result 
{that is a shoulder in the UPS spectrum centered at 1.8 eV\cite{DGSB08}. This is a distinct 
improvement over the} PBE result (ca. 1.1 eV), and matches well with previous DFT+$\Sigma$ calculations 
\cite{ELNK15}, as expected for weakly coupled interfaces. OT-RSH yields a work function 
change of -0.7 eV, similar to that of PBE (-0.5 eV) and experiment (-0.5 eV, Ref. 
\citenum{DGSB08}). {For the value of the work function, OT-RSH yields 4.9 eV, in good agreement with PBE and experiment (4.8 eV).}

\subsubsection{BDA on Au(111)}
BDA adsorbed on Au(111) has attracted {considerable} attention in both experimental 
\cite{DKCV10,HZTE13} and {theoretical\cite{TDQB11,BTNK11,LTCG12,LRLC16} studies}. However, 
the adsorption geometry {has only been} revealed recently {with} scanning tunneling 
microscopy measurements \cite{HZTE13}, {which suggest that} BDA molecules form self-assembled 
linear chains on Au(111). Ref. \citenum{LRLC16} showed that the linear-chain 
phase of BDA {is energetically favored over} isolated monomer phases used in previous 
calculations \cite{TDQB11}. {In} Ref. \citenum{LRLC16}, standard DFT+$\Sigma$ calculations 
{were performed} to correct {the} PBE PDOS, without altering the PDOS lineshape. {Here}, we 
use the same geometry as in Ref. \citenum{LRLC16}, {but calculate the PDOS of the interface 
self-consistently with our new OT-RSH approach. } We use 3 layers of Au(111) as the 
substrate, and a $4\times6\times1$ k-mesh ($2\times3\times1$ for the Fock exchange contribution). 

{The tuning approach} for the gas-phase molecule, using Eq. \eqref{tune}, yields $
\gamma=0.23$ bohr$^{-1}$ and HOMO at -7.0 eV. {As specified above,} the image plane of 
Au(111) is at 0.9 \AA~above the surface. The average distance between the molecule and the 
surface is {determined to be} 3.66 \AA. Therefore the polarization due to the substrate is 
1.3 eV, based on the image-charge model. Ref. \citenum{LRLC16} {showed} that for the 
linear-chain structure, intermolecular polarization is non-negligible, {an effect for the 
HOMO resonance that amounts to} 0.3 eV. Therefore, when choosing $\beta$ using Eq. \eqref{chooseb}, we use 
$P=1.6$ eV {as the target value}, {which requires} $\beta=0.19$. {We note in passing that for the
purpose of tuning $\beta$, we directly take the intermolecular polarization of 0.3 eV from Ref.
\citenum{LRLC16}, without attempting to determine this value from DFT or from classical electrostatic
calculations (see Refs. \citenum{RHAZ08,N10,SB12,RSJB13,V15} for such examples).}

\begin{figure}
\begin{center}
\includegraphics[width=3.5in]{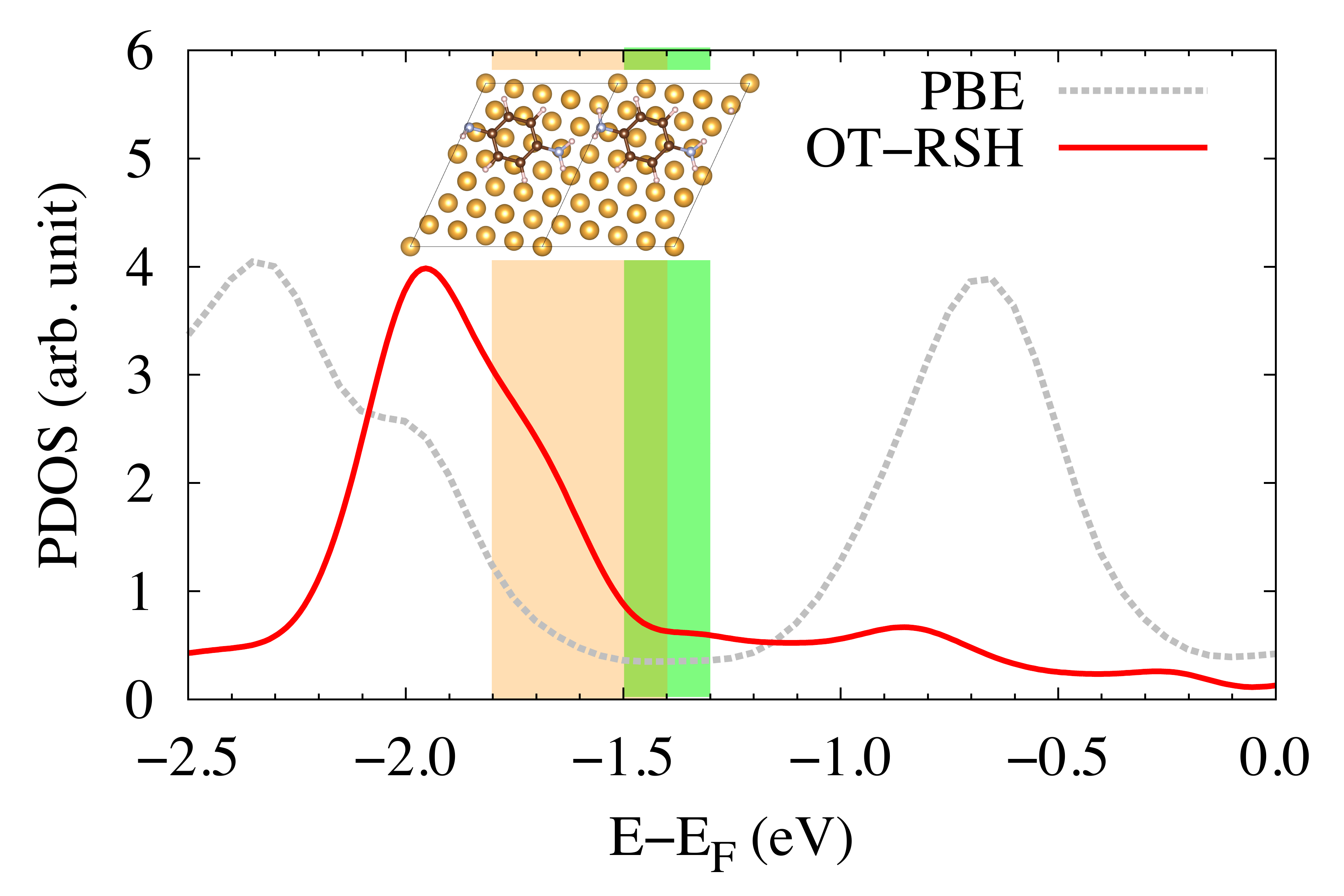}
\caption{PDOS of linear-chain phase of BDA on Au(111), whose structure is shown in the 
inset, as obtained 
using PBE {(grey dashed line)} and OT-RSH {(red line)}, compared with results from UPS and 
x-ray photoemission spectroscopy measurements 
(green and orange shaded {areas}, respectively, {the width of which represent the experimental uncertainty) from Ref.\ \citenum{DKCV10}.}}
\label{f:bda}
\end{center}
\end{figure}

Fig.\ \ref{f:bda} shows a comparison of our results to experiment\cite{DKCV10}. {As 
can be seen, the} OT-RSH HOMO resonance is at about 1.9 eV below $E_{\rm F}
$, in much better agreement with experimental results than PBE, which underestimates the 
level alignment by about 1 eV. We note that in Ref. \citenum{LRLC16}, DFT+$\Sigma$ {was 
found to} place the HOMO at 1.3 eV below $E_{\rm F}$, different from the OT-RSH result in 
this paper. This is because, firstly, Ref. \citenum{LRLC16} used {an} image plane position 
of 1.47 \AA~for Au(111), {as determined classically from linear response} 
\cite{LN93,TDQB11}, {which increases the} polarization by 0.3 eV {compared to the one 
determined here}. In addition, hybridization of molecular orbitals with metallic states can 
shift the eigenvalues when taken into account self-consistently, as compared to when using a 
post-processing approach. {Importantly}, the DFT+$\Sigma$ 
calculations in Ref. \citenum{LRLC16} only rigidly {shifts the} PBE PDOS, leaving the lineshape unchanged, in 
contrast to our OT-RSH approach with its inherent self-consistency when solving the GKS 
equation. Accordingly, in Fig.\ \ref{f:bda} one can see that OT-RSH yields a slightly {more}
asymmetric PDOS lineshape, compared to PBE, indicating enhanced hybridization with Au $d$-states at 
lower energy. Lastly, OT-RSH and PBE again yield about the same work function change (-1.5 
eV) {and the same absolute value for the work function (3.8 eV)}.

\subsubsection{AntSe on Au(111)}
\label{sec:antse}

{AntSe is known to form an upright-standing self-assembled monolayer (SAM) on 
Au(111)\cite{B08}, and is thus different from the other systems studied here. Adsorption of 
AntSe on Au(111) can be assumed to proceed through cleavage of} hydrogen atoms and 
{formation of} a covalent Au-Se bond, i.e., it is a chemisorbed system. {AntSe on Au(111) 
has been studied in detail, both experimentally and theoretically, in} Ref. 
\citenum{TRHR10}; the structure we use here was identified as the most likely one in that 
study{, which compared theory results} to experimental data from scanning tunneling microscopy. In our 
calculations, 4 layers of Au(111) {serve as the slab}, and a $8\times4\times1$ k-mesh 
($4\times2\times1$ for the Fock exchange contribution) {is used}. 

{The OT-RSH procedure for this case starts with identifying a suitable} gas-phase reference 
{system to determine the optimal RSH parameters. As mentioned above, we assume cleavage of 
hydrogen atoms upon adsorption, and thus use a H-terminated AntSe molecule (hydrogen 
optimized using PBE in the gas phase with other atoms fixed) as the reference.} {The tuning 
approach for this gas-phase molecule yields an optimal $\gamma=0.17$ bohr$^{-1}$} and HOMO 
energy {of} -7.1 eV. {As mentioned before,} {the location of the} Au(111) image plane is 
0.9 \AA~above the surface, and the average distance between the molecule and surface is 
7.06 \AA. {This gives} an image-charge energy {of} 0.6 eV. Similar to the BDA/Au(111) 
discussed above, due to the small intermolecular distances within the SAM, we expect a 
non-negligible polarization due to other molecules, in addition to the polarization due to 
the metal surface.\cite{S81,NHL06,RSJB13} {As mentioned above, we do not 
attempt to calculate this contribution from DFT alone.} To nevertheless determine this 
additional intermolecular 
polarization energy that is responsible for gap difference between an isolated molecule and 
a SAM, we carry out $GW$ calculations (see Ref. \citenum{gwnote} for technical details) using the BerkeleyGW \cite{BerkeleyGW} package for 
both the H-terminated gas-phase molecule and the SAM without the metal substrate. We find 
that the gap is reduced by 1.2 eV in the SAM compared to the gas-phase molecule. However, 
the convergence of the absolute value of $GW$ quasiparticle energies with respect to the 
vacuum level is very slow and computationally demanding. Furthermore, the dipole moment of 
AntSe complicates defining a unique vacuum level for the SAM. We therefore make the 
assumption that this gap change equals $2P$, which is reasonable because the shapes of the 
HOMO and LUMO are similar. It then follows that the HOMO renormalizes in the SAM by 0.6 eV, 
with respect to the gas-phase molecule. The total renormalization of the HOMO, summing 
contributions from the substrate and from other molecules, is therefore 1.2 eV, which gives 
rise to $\beta=0.21$ for the interface, according to Eq. \eqref{chooseb}.

\begin{figure}
\begin{center}
\includegraphics[width=3.5in]{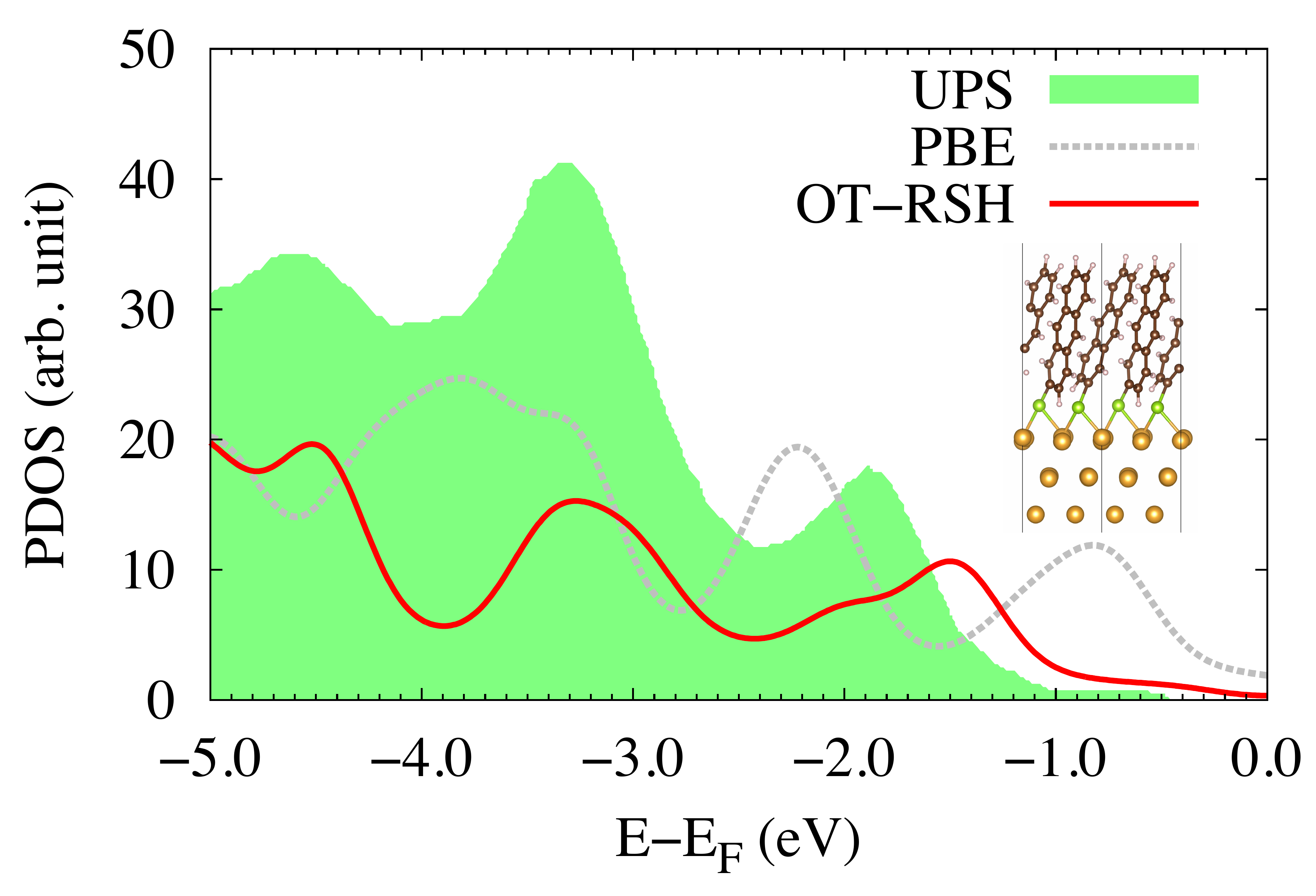}
\caption{PDOS of AntSe adsorbed on Au(111), whose structure is shown in the inset, as obtained 
using PBE 
{(grey dashed line)} and OT-RSH {(red line)}, compared with {results from UPS measurements} 
(green) taken from Ref. \citenum{TRHR10}.}
\label{f:antse}
\end{center}
\end{figure}

Fig.\ \ref{f:antse} shows our results compared to experimental data \cite{TRHR10} for this 
system. OT-RSH predicts a HOMO resonance at 1.5 eV below $E_{\rm F}$, in much better agreement with 
the experimental result (1.9 eV) than PBE, which places the HOMO resonance at 0.8 eV below $E_{\rm F}
$. Furthermore, in the OT-RSH PDOS the consecutive features {at} higher binding energies 
(lower in energy in Fig.\ \ref{f:antse}) also match the UPS data very well, in sharp 
contrast to PBE. The good agreement of OT-RSH shows that the gas-phase reference is 
physically reasonable, meaning that the HOMO orbital is nearly unaffected by the 
chemisorption, which makes sense given that it is delocalized over the backbone and not 
localized on the Se linker. Lastly, both OT-RSH and PBE yield {excellent agreement with
experiment for both the work function change (about 1.3 eV) and its absolute value (3.9 eV in PBE and experiment and 4.0 eV in OT-RSH).}

\subsection{Systems with non-negligible charge transfer}
\label{sec42}
In this section we present two systems involving the Ag(111) substrate with the adsorbates 
PTCDA and 1,4,5,8-naphthalene-tetracarboxylic dianhydride (NTCDA). Because Ag(111) has a 
lower work function than Au(111), the coupling between the molecule and substrate is 
stronger and {signatures in UPS that energetically close to the Fermi level are ascribed to 
the (formerly unoccupied) LUMO of the molecule}. This signals non-negligible 
charge transfer between the metal and the molecule, {for which} standard DFT+$\Sigma$ {and 
other non-self-consistent approaches} {run into difficulties} because {the weak-coupling} 
assumption of DFT+$\Sigma$ is violated. In fact, a naive application 
of DFT+$\Sigma$ {is challenging for} the pinned LUMO, which is of course unoccupied in the 
gas phase, but at least partially occupied when the molecule adsorbs on the metal surface. 
We will show {in the following} that our proposed OT-RSH approach also {performs} 
well for such cases.

\subsubsection{PTCDA on Ag(111)}
PTCDA interacts more strongly  {with the} Ag(111) than {with the} Au(111) surface and, {as 
observed in UPS and transport experiments,} features a LUMO peak pinned a $E_{\rm F}$, see, 
e.g., {Refs. \citenum{Z06,K06,HBLS07,T07,R07,DGSB08,R09,Z10, HAMR13,SLWR14} for detailed 
experimental and theoretical discussions}. This charge 
transfer is already captured by LDA/GGA functionals \cite{R07,R09}. 
However, {we would expect} the LDA/GGA HOMO {resonance position (i.e., 
the second highest peak in energy in UPS) to be higher in energy than the physical one}, just 
as {for} the cases discussed {above}. The OT-RSH {method} proposed in this work, as we will 
show below, corrects this quantitative error {for the HOMO energy alignment}, and at the same time 
maintains {a correct description of} the LUMO pinning {effect}. The geometry {we use is 
taken from Ref. \citenum{RLZS12} and has 3 layers of Ag(111) as the substrate, and we 
employ} a $2\times2\times1$ ($1\times1\times1$ for the Fock exchange contribution) k-mesh.

\begin{figure}
\begin{center}
\includegraphics[width=3.5in]{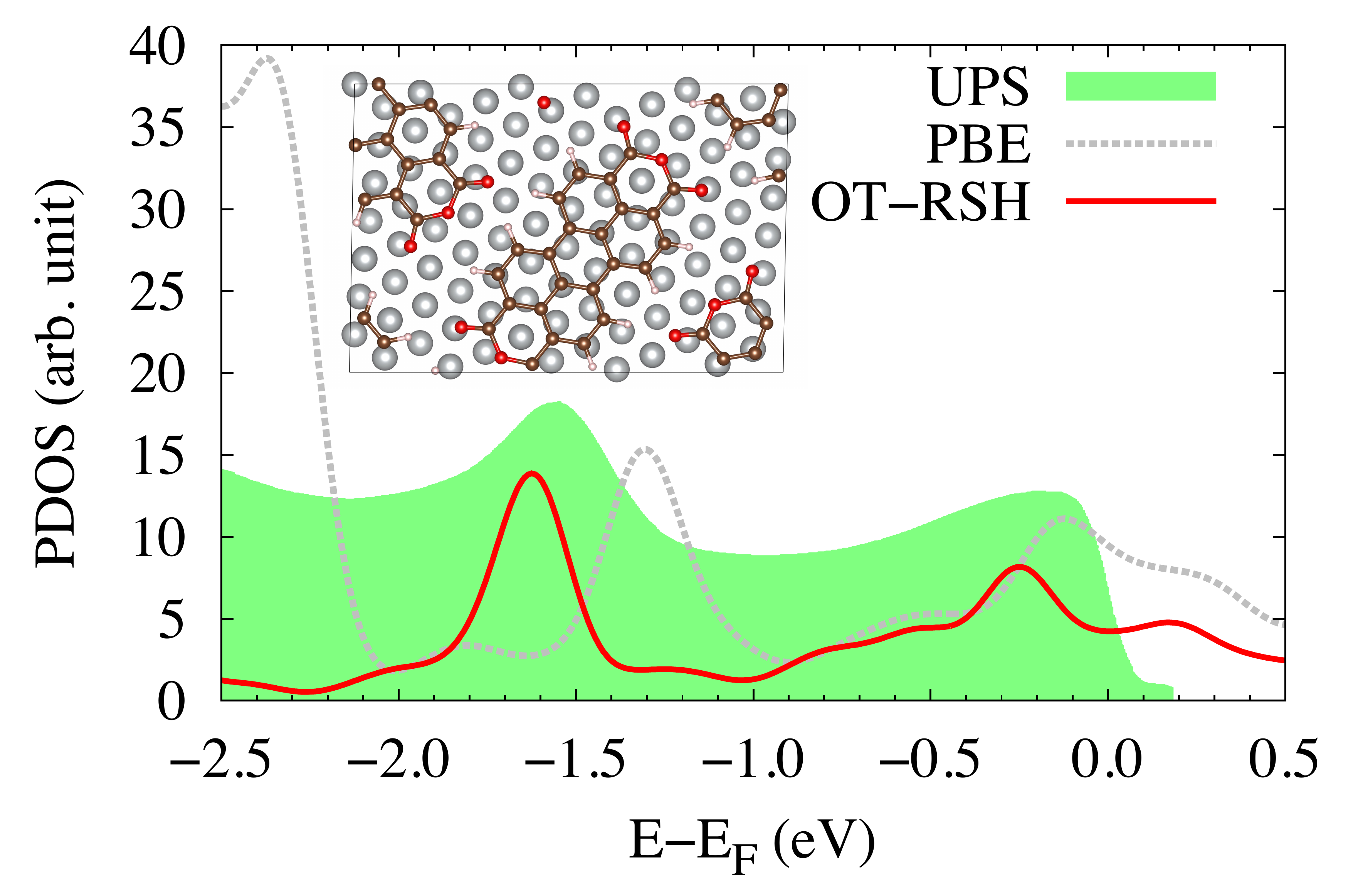}
\caption{PDOS of PTCDA adsorbed on Ag(111), whose structure is shown in the inset, as obtained 
using PBE 
{(grey dashed line)} and OT-RSH {(red line)}, compared with {results from UPS measurements} 
(green) taken from Ref. \citenum{SLWR14}.}
\label{f:ptcdag}
\end{center}
\end{figure}

{For this system, the OT-RSH procedure is the same as discussed above:} tuning of gas-phase 
PTCDA yields {an} optimal $\gamma=0.15$ bohr$^{-1}$ and a HOMO energy of -8.1 eV. {Note that the 
gas-phase HOMO energy and optimal $\gamma$ of PTCDA is slightly different from the one 
reported above for the PTCDA/Au(111) case, as we use the molecular coordinates optimized 
for the interface system, which change due to the different interactions with the Au or Ag 
substrate\cite{R07}.} For Ag(111), the image plane {position} is {determined to be} 1.0 
\AA~above the surface. The average distance between PTCDA and the Ag surface is 2.87 \AA, 
giving rise to an image-charge energy of 1.9 eV. {When we tune $\beta$ using Eq. 
\eqref{chooseb} for this adsorbate, a complication arises:} with the physically lowest 
possible $\beta$, {as specified above ($\beta=-\alpha=-0.2$)}, the HOMO only increases by 
1.7 eV. This is not 
enough to accommodate the image-charge energy, {with the remaining difference being 0.2 eV}. 
This large nonlocal surface polarization is again a sign of strong coupling. In such 
cases, we simply use $\beta=-0.2$ {to perform the OT-RSH calculation of the interface}.

Fig.\ \ref{f:ptcdag} shows {our calculated results} for PTCDA adsorbed on Ag(111) compared 
{to data from UPS} measurements \cite{SLWR14}. {The OT-RSH PDOS shows} the {important} 
feature of LUMO pinning at $E_{\rm F}$, and on top of that corrects the PBE HOMO resonance 
energy, placing it at 1.6 eV below $E_{\rm F}$ and achieving better agreement with 
experiment. OT-RSH and PBE results for the work function change are both very small and 
close to zero, comparable to experimental results. 
{Both OT-RSH and PBE yield a work function of 4.6 eV, in good agreement with experiment.} 
These findings are highly encouraging 
and show that our OT-RSH method captures all the relevant and highly non-trivial effects at 
this interface.

\subsubsection{NTCDA on Ag(111)}
\label{perthyb}
Another {interesting system that shows strong interaction} is NTCDA adsorbed on Ag(111). 
Similar to PTCDA on Ag(111), it also features a LUMO peak at $E_{\rm F}$, as evidenced in 
UPS measurements\cite{BFSB07,SKZZ10,ZHKB12}. {We relax the NTCDA-Ag(111) geometry} with {the PBE+vdW$^{\rm surf}$} 
{method} \cite{RLZS12} as implemented in VASP, using 3 layers of Ag(111) and 
a $4\times4$ surface unit cell as the substrate. The level alignment calculations are 
carried out using a $4\times4\times1$ k-mesh ($2\times2\times1$ for the Fock exchange 
contribution). {From our tuning,} the optimal $\gamma$ for the gas-phase molecule is 0.19 bohr$^{-1}$, 
according to Eq. \eqref{tune}, and its HOMO energy is -9.7 eV. For Ag(111), the 
image-plane {position} is at 1.0 \AA~above {the} surface{, as determined above}. The 
average distance between the {optimized} molecule and the surface is 2.88 \AA, yielding an 
image-charge energy {of} 1.9 eV. {This requires} $\beta=-0.082$ for the interface, 
according to Eq. \eqref{chooseb}.

\begin{figure}
\begin{center}
\includegraphics[width=3.5in]{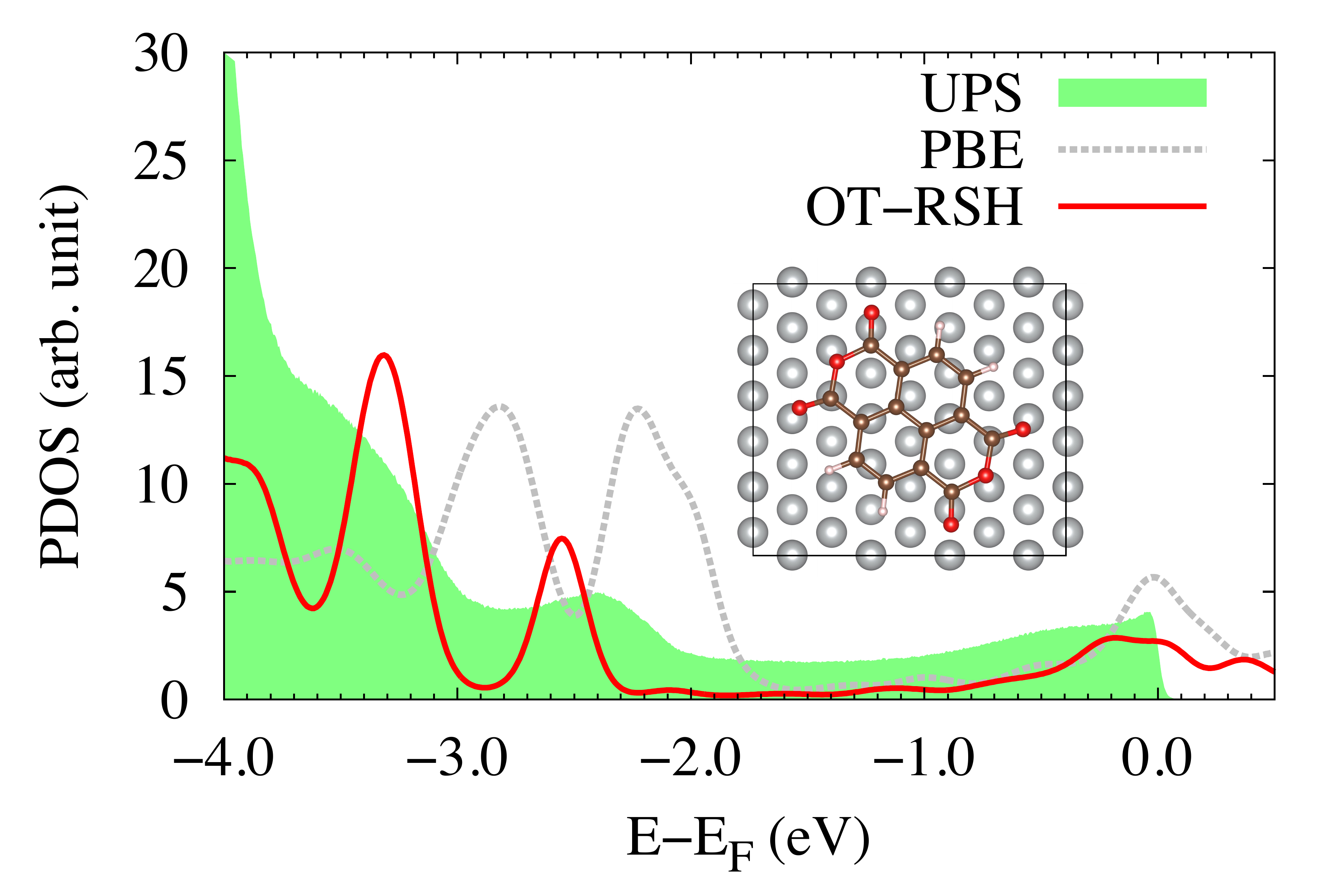}
\caption{PDOS of NTCDA adsorbed on Ag(111), whose structure is shown in the inset, as obtained 
using PBE 
{(grey dashed line)} and OT-RSH {(red line)}, compared with {results from UPS measurements} 
(green) taken from Ref. \citenum{SKZZ10}.}
\label{f:ntcda}
\end{center}
\end{figure}

Fig.\ \ref{f:ntcda} {shows our results and UPS data of} Ref. \citenum{SKZZ10}. {Compared to 
the experimental results, OT-RSH correctly depicts the pinned LUMO, the HOMO resonance, and 
places the HOMO-1} where UPS shows a shoulder, already around 3.5 eV below $E_{\rm F}$. PBE 
correctly shows a LUMO pinned at $E_{\rm F}$, {describes the HOMO reasonably well, but 
misplaces the} HOMO-1 resonance (by 0.7 eV) with respect to results from OT-RSH and experiment. {This 
strongly suggests that} OT-RSH is not only accurate for the {frontier resonances}, but 
{actually maintains predictive power} within several eVs from $E_{\rm F}$,\cite{RSGA12,EWRS14} owing to the 
nonlocal exchange operator \cite{K10}. {OT-RSH and PBE yield similar work function
values (4.7 eV and 4.8 eV, respectively).} Regarding the work function change, OT-RSH 
and PBE differ slightly more (by 0.4 eV and have different signs) in this case. However, we 
are not aware of experimental data for the work function change in this case, and our results suggest this would be a measurement of interest.

\begin{table*}
\centering
\caption{Summary of the OT-RSH and PBE results and experimental literature data for the 
interface systems studied in this work. $E_{\rm F}-E_{\rm HOMO}$ is the level alignment for 
the HOMO resonance, and $\Delta \Phi$ is the work function change compared to the clean 
metal surface. {$\Phi$ is the resulting work function of the interface. Where we are not aware of an experimental $\Delta \Phi$ or $\Phi$ value, an ``x'' sign is used.} $\gamma$ denotes the optimally-tuned range-separation parameter based on 
Eq. \eqref{tune} and is in bohr$^{-1}$. $\varepsilon_{\rm HOMO}^{\rm OT-RSH}$ is the gas-phase 
OT-RSH HOMO energy. $z_0$ is the DFT-determined image-plane position for the metal 
surface. $P$ is the surface polarization calculated using an image-charge model. For 
systems with small surface unit cells, such as BDA/Au(111) and AntSe/Au(111), $P$ also 
includes polarization contribution due to other molecules in the molecular layer. $\beta$ 
is tuned according to Eq. \eqref{chooseb} and is used for the calculation of the interface. 
The first four systems show negligible charge transfer and are discussed in Sec. 
\ref{sec41}. The last two systems show significant charge transfer and are discussed in 
Sec. \ref{sec42}.}
\resizebox{\textwidth}{!}
{
\begin{tabular}{c|ccc|ccc|ccccc|c}
\hline\hline
 & \multicolumn{3}{c|}{$E_{\rm F}-E_{\rm HOMO}$(eV)} & \multicolumn{3}{c|}{$\Delta \Phi$ {[$\Phi$]}
(eV)} & \multirow{2}{*}{$\gamma$($a_0^{-1}$)} & \multirow{2}{*}{$\varepsilon_{\rm HOMO}
^{\rm OT-RSH}$(eV)} & \multirow{2}{*}{$z_0$(\AA)} & \multirow{2}{*}{$P$(eV)} & \multirow{2}
{*}{$\beta$} & \multirow{2}{*}{comment} \\
 & PBE & OT-RSH & Expt. & PBE & OT-RSH & Expt. & & & & & & \\
\hline
Benzene/Al(111) & 3.1 & 4.4 & 4.0 \cite{DMBN00} & -0.3 {[3.8]} & -0.3 {[3.8]} & -0.2 {[x]} \cite{DMBN00} & 0.24 & 
-9.4 & 1.1 & 1.7 & 0.20 & weak physisorption \\ 
PTCDA/Au(111) & 1.1 & 1.4 & 1.6-2.0 \cite{DGSB08} & -0.5 {[4.8]} & -0.7 {[4.9]} & -0.5 {[4.8]} \cite{DGSB08} & 0.16 
& -8.2 & 0.9 & 1.6 & -0.10 & weak physisorption \\
BDA/Au(111) & 0.7 & 1.9 & 1.3-1.8 \cite{DKCV10} & -1.5 {[3.8]} & -1.6 {[3.8]} & x {[x]} & 0.23 & -7.0 & 0.9 & 1.6 
& 0.19 & stronger hybridization \\
AntSe/Au(111) & 0.8 & 1.5 & 1.9 \cite{TRHR10} & -1.4 {[3.9]} & -1.5 {[4.0]} & -1.3 {[3.9]} \cite{TRHR10} & 0.17 & 
-7.1 & 0.9 & 1.2 & 0.21 & with covalent bond \\ 
\hline
PTCDA/Ag(111) & 1.3 & 1.6 & 1.5 \cite{SLWR14} & +0.1 {[4.6]} & 0.0 {[4.6]} & -0.1 {[4.8]} \cite{DGSB08},0.1 {[4.9]}\cite{Z06} & 0.15 & 
-8.1 & 1.0 & 1.9 & -0.20 & Fermi level pinning \\
NTCDA/Ag(111) & 2.2 & 2.5 & 2.4 \cite{SKZZ10} & +0.2 {[4.8]} & -0.2 {[4.7]} & x {[x]} & 0.19 & -9.7 & 1.0 & 1.9 & 
-0.082 & Fermi level pinning \\
\hline
\end{tabular}
}
\label{tab:res}
\end{table*}

\section{Discussion}
\label{sec4}

{In this paper, we have proposed an approach for accurate calculation of level alignments 
at molecule-metal interfaces based on the OT-RSH functional, and tested it for a set of 
well-studied and complex systems. Our results, summarized in Table \ref{tab:res}, show that 
this method} is successful in quantitative predictions of energy level alignment. To 
understand its success, we refer again to Fig.\ \ref{rshcartoon} and compare it to Fig.\ 
\ref{alignment}: The essential idea of our OT-RSH {approach} is to tune the parameter $\beta$ {and adapt 
the long-range Fock exchange} to capture the screening due to the metal surface. The amount 
of screening and the choice of $\beta$ are determined by $P$, {given} in Eq. 
\eqref{chooseb}. In Fig.\ \ref{rshcartoon}, the {red} lines on the right demonstrate the {attempted} 
gas-phase HOMO/LUMO energy with the tuned $\beta$ value, and the red curve on the left shows the 
resulting PDOS of the molecule in the interface. {In Eq. \eqref{chooseb}, for the systems studied here the left hand side is almost linear in $\beta$, which means that the tuned $\beta$ value, as well as the resulting level alignment at the interface, are not sensitive to small variations of $P$.}

{With our method we set out to capture physics inherent to $GW$ calculations but absent 
from currently available DFT approaches for interfaces, namely a sensitivity of XC effects 
to the dielectric environment specific to molecule-metal interfaces. The {uneven} 
performance of standard DFT approaches for interfaces, i.e., the success in describing work 
functions and interface dipoles but the failures in predicting level alignments, can be 
understood from the ``nearsightedness principle'' of many-electron systems\cite{K96}. It 
states that many important static physical quantities, including the density, only depend 
on changes in the potential at nearby points. This explains the success of local and semi-local 
XC functionals in predicting the density of molecule-metal systems and highly 
relevant quantities that are directly determined from it, notably total energies, work 
functions, and interface dipoles. However, long-range Coulomb interactions are an exception 
to this principle. They are important for molecule-metal interfaces, especially considering 
the renormalization of molecular resonances due to the metal surface.} Because of the long-range nature of the phenomenon, a 
standard hybrid functional without range separation, or a range-separated hybrid functional
without long-range Fock exchange, cannot in general capture the image-charge effect accurately. Therefore, we choose to 
adapt the long-range part of the XC potential by tuning $\beta$ to 
capture $P$, thereby implicitly including the otherwise missing distance-dependent 
screening in our approach, which is crucial for its success. At the same time, we 
deliberately do not modify the short-range part of the XC potential (controlled 
by $\alpha$ and $\gamma$) when going from the gas phase to the surface. Importantly, 
orbital energies and resonances are expected to respond to $\beta$ tuning, but the density 
and quantities that are directly determined from it are expected to be mostly unchanged. 
This reasoning is strongly supported by 
our results of level alignments and work function changes. Our scheme therefore provides a 
superior GKS approach that can predict both energy level alignments and work function 
changes with high accuracy.

As we have stressed {throughout this paper}, one advantage of {our fully self-consistent 
approach, as} compared to DFT+$\Sigma$ {and other non-self-consistent schemes,} is that it 
calculates {and fully captures} the hybridization between the molecule and the metal. 
Therefore, it can alter the PBE PDOS lineshape and {is also applicable} to systems with 
stronger charge transfer and Fermi level pinning. In addition, since it is based on GKS and 
{includes a fraction of short-range Fock exchange,} it can also yield {improved relative 
orbital spacing,} as shown in the examples of AntSe/Au(111) and NTCDA/Ag(111), as well as 
improved orbital ordering, as {was} demonstrated previously for other systems
\cite{D06,R11,P11,H14,EWRS14,LWYA14}.

In spite of its success, we also would like to point out that the method {proposed here} is 
not a panacea. In particular, we focused {here} on the 
level alignment and PDOS of the molecule, and did not discuss how the specific choice of $
\alpha,\beta,$ and $\gamma$ {may} affect {specifics in the electronic structure} of the 
metal substrate. {In this context, we would like to mention that} full Fock exchange 
often artificially opens a gap for metals \cite{mahan}, and {that} many hybrid functionals 
perform worse than {local and} semi-local functionals for metals \cite{PMK07}. We expect 
{that similar situations may occur when one has to choose} a large $\beta$ {value} 
following the tuning procedure. {Coincidentally,} this is not the case for the systems 
studied in this work. The largest $\beta$ used in this work is about 0.2, and with this 
amount of long-range Fock exchange the PDOS of the metal substrate stays qualitatively 
correct. {Quantitatively, even for a rather small amount of Fock exchange, the theoretical 
description of certain properties of real metals may suffer\cite{BTNK11}.}

Another limitation is {that for PTCDA/Ag(111), and perhaps also other systems, with the 
physically lowest possible value for $\beta$, i.e., $\beta=-\alpha=-0.2$} the left hand 
side is smaller than the right hand side of Eq. \eqref{chooseb}. In other words, {we cannot 
accommodate the full magnitude of $P$ by tuning $\beta$}. {For PTCDA/Ag(111), the remaining 
difference was} 
small. However, one {may} imagine 
{molecule-metal} systems where the deficit is {somewhat} larger, and using $\beta=-0.2$ 
would not be enough. These scenarios {are more likely to occur} for more strongly coupled 
interfaces, such as pentacene adsorbed on Ag or Cu \cite{ULRK14}. In those cases, the 
currently proposed OT-RSH approach {may} not work as well as for the systems shown in this 
paper.

Lastly, we {have used a} DFT-based image-charge model {to approximate} the {otherwise 
rather} complicated change in the Coulomb screening induced by the metal surface. As far as 
this particular approximation is concerned, the approach proposed in this work shares the 
same advantages and disadvantages as the ``standard'' DFT+$\Sigma$ method: The image-charge 
interaction mimics static (frequency-independent) polarization, and misses dynamical 
(frequency-dependent) surface polarization effects and the polarization of the molecule due 
to the metal, which is a higher order effect. {Moreover, for cases of non-negligible 
intermolecular polarization such as polarizable SAMs, so far we did not attempt to compute 
this solely from DFT, but for this proof-of-principle study relied on $GW$ results.}

One way of understanding the above limitations is to realize that molecule-metal interfaces 
are {indeed} strongly heterogeneous {systems} in the sense that the screening perpendicular 
to the surface is strongly distance-dependent and very different from the screening 
parallel to it. In the OT-RSH approach proposed {here}, we use \emph{scalar} parameters, $
\alpha,\beta,$ and $\gamma$, which may not be enough for complicated cases. Work along 
the lines of ``local hybrids'' \cite{JSE03}, i.e., spatially dependent parameters $
\alpha(\mathbf{r}),\beta(\mathbf{r}),$ or even $\gamma(\mathbf{r})$, {as well as} density-based 
mixing parameters \cite{MVOR11}, may describe the heterogeneous environment more 
accurately. But the construction of such functionals and their implementations may {prove 
to} be much more difficult.

\section{Conclusion}
\label{sec5}
In this work, we developed an approach to calculate the energy level alignment at 
molecule-metal interfaces with good accuracy, based on an OT-RSH functional. The essential idea is to start 
with an accurate electronic structure description in the gas phase and capture the nonlocal 
surface polarization due to the metal by screening the long-range Fock exchange. We 
proposed a non-empirical way to tune the long-range Fock exchange per molecule-metal pair 
based on an image-charge model, and implemented this {approach} in a plane-wave code. 
Results from our fully self-consistent approach for several prototypical, challenging 
molecule-metal interfaces are in quantitative agreement with experiments, for both the 
level alignments and work function changes. Keeping its remaining limitations as discussed 
in this paper in mind, we believe that our OT-RSH approach paves the way for accurate and 
reliable predictions of energetics and level alignments at heterogeneous interfaces, 
especially those related to energy conversion and molecular electronics. Finally, we note 
that the efficiency of these calculations strongly depends on new algorithm developments 
for computing nonlocal Fock exchange, such as the recently proposed adaptively compressed 
exchange operator method \cite{L16}.

\section{Acknowledgement}

We thank Achim Sch\"{o}ll for providing us the experimental data for NTCDA adsorbed on 
Ag(111) that is published in Ref. \citenum{SKZZ10}. Work in Berkeley was supported by the 
U.S. Department of Energy, Office of Basic Energy Sciences, Division of Materials Sciences 
and Engineering under Contract No. DE-AC02-05CH11231. Work performed at the Molecular 
Foundry was also supported by the Office of Science, Office of Basic Energy Sciences, of 
the U.S. Department of Energy under the same contract number. Work in Rehovoth was 
supported by the European Research Council, the Israel Science Foundation, the United 
States-Israel Binational Science Foundation, the Wolfson Foundation, 
the Austrian Science Fund (FWF):J3608-N20, and the Molecular Foundry. We thank 
the National Energy Research Scientific Computing center for computational resources.

\singlespacing
\bibliography{lit_interfaceRSH}

\begin{thebibliography}{153}
\bibitem [{mix()}]{mixpp}%
  \BibitemOpen
  \href@noop {} {}\bibinfo {howpublished} {We have checked the effect of mixing
  these two types of pseudopotentials on the level alignment, using the system
  of 2,3,5,6-tetrafluoro-7,7,8,8-tetracyano-quinodimethane (F4TCNQ) adsorbed on
  Ag(111) surface. The calculation is performed using HSE functional and the
  resulting level alignment is in quantitative agreement with both PAW
  calculations in VASP and the published results in Ref. \citenum{HAMR13} using
  the same functional.}\BibitemShut {Stop}%
\bibitem [{gwn()}]{gwnote}%
  \BibitemOpen
  \href@noop {} {}\bibinfo {howpublished} {In the $GW$ calculations, 1500 bands
  and an energy cutoff of 10 Ry are used to calculate the polarizability. The
  dielectric function is calculated using the generalized plasmon pole model. A
  box truncation and a single q-point ($\Gamma$) are used for the molecule. A
  slab truncation and a $8\times 4 \times 1$ q-mesh are used for the monolayer.
  Static remainder is used in the evaluation of self-energy
  correction.}\BibitemShut {Stop}%
\end{thebibliography}
\end{document}